\documentclass[twocolumn,showpacs,preprintnumbers,amsmath,amssymb]{revtex4}
\usepackage{graphicx}
\usepackage{dcolumn}
\usepackage{bm}
\begin{document}

\title{INFORMATION CAPACITY OF BIOLOGICAL MACROMOLECULAE RELOADED}

\author{Michael G.Sadovsky}
\affiliation{Institute of biophysics of SD of RAS;\\ 660036 Russia, Krasnoyarsk,
Akademgorodok.} \email{msad@icm.krasn.ru}

\begin{abstract}
Information capacity of a symbol sequence is a measure of the unexpectedness of a
continuation of given string of symbols. Continuation of a string is determined through
the maximum entropy of the reconstructed frequency dictionary; the capacity, in turn, is
determined through the calculation of mutual entropy of a real frequency dictionary of a
sequence with respect to the reconstructed one. The capacity does not depend on the
length of strings in a dictionary. The capacity calculated for various genomes exhibits a
multi-minima pattern reflecting an order observed within a sequence.
\end{abstract}


\maketitle

\section{\label{introd}Introduction}
The analysis of statistical patterns in completely sequenced genomes is of great
interest. The correlations observed within these latter reflect some biological features
of primary structures \cite{r1,r2,r3,r4}. In particular, the sequence periodicity of $3$
base pairs (bp) indicates the presence of protein coding regions in a genome; more
exactly, non-coding regions are invariant against the frame shift of the codon pattern,
while the coding ones lack these invariance \cite{z1,z1-1,z2}.

A complexity of patterns observed in a genetic sequence may vary significantly. The
complexity itself is a matter of interests of mathematicians, biologists and
biophysicists \cite{z3,z4,z5,z6,a1,a2,a3,a4}. Screening a genome with respect to a
complexity of different fragments of that latter, a student my find various biologically
important peculiarities in a nucleotide sequence. Information capacity measurements
bring, in turn, a new knowledge towards the genetic entities. Here we present a new
approach to determine the information capacity of a symbol sequence, with applications to
genetic sequences.

To begin with, it should be stressed, that a symbol sequence has zero information
content, zero information capacity, and zero redundancy, being a finite object. To
discuss all these issues with particular respect to the genetic entities, one must change
a finite sequence for a frequency dictionary of short strings. Such transformation makes
a student change a consideration of a finite sequence for the ensemble of (infinite)
sequences, which yield the same distribution of frequency of short strings, as a given
dictionary determined over a finite sequence; further, we shall no more mention this
difference, provided that no misunderstanding occurs. If any, special remarks would be
done.

Physically, DNA sequence is a polymer molecule, which could be considered as a symbol
sequence from the four-letter alphabet $\aleph = \{\mathsf{A}, \mathsf{C}, \mathsf{G},
\mathsf{T}\}$, where $\mathsf{A}$ refers to adenine, $\mathsf{C}$ refers to cytosine,
$\mathsf{G}$ refers to guanine, and $\mathsf{T}$ refers to thymine. Let $N$ be the length
of the sequence, i.e. the number of symbols in it. Further, we shall consider the
continuous sequences only; a consideration of unbound sequences is possible, while brings
no new comprehension, but the technical problems \cite{s4,s5}.

Any continuous string of the length $q$, $1 \leq q \leq N$ observed within a sequence
makes a \textbf{word} $\omega = \nu_1\nu_2\nu_3\ldots \nu_{q-1}\nu_q$ (of the length
$q$); here $\nu_j \in \{\mathsf{A}, \mathsf{C}, \mathsf{G}, \mathsf{T}\}$. A set of all
the words (of the given length $q$) observed at the sequence makes the \textbf{support}
of that former (or $q$--support, if indication of the length is necessary). Providing
each element of a support (i.e., each word $\omega$) with the number $n_{\omega}$ of
copies of that latter, one gets the dictionary $\mathfrak{F}_q$ of the sequence (of the
thickness $q$). The dictionary $\mathfrak{F}_q$ is a finite object, as well. Changing the
number of copies $n_{\omega}$ for frequency: $$f_{\omega} = \frac{n_{\omega}}{N} \;,$$
one gets a \textbf{frequency dictionary} $W_q$ (of the thickness $q$).

Such definition of frequency requires a connection of a sequence into a ring; the
motivation behind such transformation is simple and obvious. Any dictionary of the
thickness $q$ could easily be transformed into a thinner dictionary $q-1$. To get the
dictionary $W_{q-1}$, one must sum up the frequency of words differing in the first, or
in the last symbol. Being provided over a dictionary $\mathfrak{F}_q$, these two
summations would yield two different results; the difference results from a finite
sampling. The starting $q-1$ symbols will not be accounted, if the summation would be
carried over the first symbol; reciprocally, the ending $q-1$ symbols will be lost, if
the summation would be carried out over the last symbol. A connection of a sequence into
a ring eliminates the problem.

Consider a symbol sequence. Compose, then, a chain of dictionaries $W_j$ of increasing
thickness $j$:
\begin{equation}\label{zep}
W_1 \leftrightarrow W_2 \leftrightarrow \ldots \leftrightarrow W_q\leftrightarrow
W_{q+1}\leftrightarrow \ldots \leftrightarrow W_N \leftrightarrow \ldots \;.
\end{equation}
Study of statistical properties of symbol sequences means an investigation of the
relations between the dictionaries within the chain. A downward transformation, i.e., the
transition from $W_j$ to $W_{j-1}$ is simple and unique. On the contrary, the upward
transformation is ambiguous, in general. A word $\omega$ may have several continuations
(not more than 4, in case of nucleotide sequences). Such ambiguity results in a positive
information capacity of the relevant frequency dictionary.

\section{\label{sec:level3}Reconstructed frequency dictionary}
An ambiguity of transformation of a thinner dictionary into a thicker one rises a
question towards the reconstruction of the dictionary. Indeed, while the downward
transformation $W_j \rightarrow W_{j-1}$ is unique, the upward transformation, in
general, generates several dictionaries. A transformation of $W_j$ into $W_{j+1}$
consists in a combination of words from the frequency dictionary $W_j$ so, that the
dictionary bearing the combined longer words yields the original frequency dictionary. In
other words, each combined set $f^{\ast}_{\nu_1\nu_2\nu_3\ldots \nu_{q-1}\nu_q\nu_{q+1}}$
of longer words must meet the constraint
\begin{eqnarray}\label{cnst}
\sum_{\nu_{q+1}} f^{\ast}_{\nu_1\nu_2\nu_3\ldots \nu_{q-1}\nu_q\nu_{q+1}} =\nonumber\\ =
\sum_{\nu_{q+1}} f^{\ast}_{\nu_{q+1}\nu_1\nu_2\nu_3\ldots \nu_{q-1}\nu_q} =\nonumber\\ =
f_{\nu_1\nu_2\nu_3\ldots \nu_{q-1}\nu_q}\; ,
\end{eqnarray}
where $f_{\nu_1\nu_2\nu_3\ldots \nu_{q-1}\nu_q}$ is the frequency of a word
$\omega=\nu_1\nu_2\nu_3\ldots \nu_{q-1}\nu_q$ from the given frequency dictionary. Linear
constraints (\ref{cnst}) eliminate a part of possible combinations of words, but an
abundance of the combinations is still great enough.

Since a set of dictionaries $\{W^{(1)}_{q+1}, W^{(2)}_{q+1}, W^{(3)}_{q+1}, \ldots,$ $
W^{(k)}_{q+1}\}$ still meet the constraint (\ref{cnst}), one has to figure out a single
one that is expected to be the reconstructed entity. Such reconstructed dictionary is
identified through the maximum entropy
\begin{equation}\label{me}
\max_{j} \left\{- \sum_{\omega^{\ast}} f^{(j)}_{\omega^{\ast}} \cdot \ln
f^{(j)}_{\omega^{\ast}} \right\}
\end{equation}
of a dictioary, where $\omega^{\ast} = \nu_1\nu_2\nu_3\ldots \nu_{q-1}\nu_q\nu_{q+1}$ is
a word meeting the linear constraint (\ref{cnst}). The dictionary $\widetilde{W}_{q+1}$
meeting the maximal principle (\ref{me}) exists always, since the set of the dictionaries
which could be combined from the given one is finite.

The frequency of words $\widetilde{\omega} \in \widetilde{W}_{q+1}$ could be figured out
explicitly, by LaGrange multiplier method \cite{n1,n2,n3,kitai}. Frequency of a word
$\omega$ of the reconstructed dictionary $\widetilde{\omega}$ is determined by the
expression
\begin{equation}\label{vosst}
\widetilde{f}_{\nu_1\nu_2\nu_3\ldots \nu_{q-1}\nu_q\nu_{q+1}} =
\frac{f_{\nu_1\nu_2\nu_3\ldots \nu_{q-1}\nu_q} \times f_{\nu_2\nu_3\ldots
\nu_{q-1}\nu_q\nu_{q+1}}}{f_{\nu_2\nu_3\ldots \nu_{q-1}\nu_q}}\;.
\end{equation}
The expression (\ref{vosst}) coincides perfectly to the Kirkwood's approximation
\cite{kirk}; an absence of the interaction via the third ``particle" makes the expression
(\ref{vosst}) here an exact solution of the problem.

Actually, the maximal entropy principle (\ref{me}) allows to reconstruct the dictionary
$\widetilde{W}_{q+l}$ for any $l>1$. Here we provide a final formula
\begin{widetext}
\begin{equation}\label{dlinno}
\widetilde{f}_{\nu_1\nu_2\nu_3\ldots \nu_{q+l-1}\nu_{q+l}} =
\frac{f_{\nu_1\nu_2\nu_3\ldots \nu_{q-1}\nu_q} \times f_{\nu_2\nu_3\nu_4\ldots
\nu_q\nu_{q+1}} \times \ldots \times f_{\nu_{l}\nu_{l+1}\nu_{l+2}\ldots
\nu_{q+l-2}\nu_{q+l-1}} \times f_{\nu_{l+1}\nu_{l+2}\nu_{l+3}\ldots
\nu_{q+l-1}\nu_{q+l}}}{f_{\nu_2\nu_3\nu_4\ldots \nu_{q-1}\nu_q} \times
f_{\nu_3\nu_4\nu_5\ldots \nu_q\nu_{q+1}} \times \ldots \times
f_{\nu_{l+2}\nu_{l+3}\nu_{l+4}\ldots \nu_{q+l-2}\nu_{q+l-1}}} \; ;
\end{equation}
\end{widetext}
see \cite{n1,n2,n3} for details.

Reconstruction of a thicker dictionary (\ref{vosst}) due to the maximum entropy principle
yields the dictionary $\widetilde{W}_{q+1}$ (or $\widetilde{W}_{q+l}$, respectively),
that bears no outer, additional information. It contains the words of the length $q+1$
(of the length $q+l$, respectively) that are the most probable continuations of the words
of the length $q$. The reconstructed dictionary $\widetilde{W}_{q+1}$ bears all the words
that occur at the dictionary $W_{q+1}$ and, maybe, some other ones. For any $q$, $q \geq
1$ $$S\left[\widetilde{W}_{q+1}\right] \geq S\left[W_{q+1}\right] \;.$$
\begin{table*}
\caption{\label{pombe}Information capacity (\ref{infcap}) determined for three
chromosomes of {\sl Schizosaccharomyces pombe} complete genome and for eleven chromosomes
of {\sl Encephalitozoon cuniculi}. $q$ is the dictionary thickness.}
\begin{ruledtabular}
\begin{tabular}{r|ccc|ccccccccccc}
\multicolumn{1}{c}{} & \multicolumn{3}{c|}{{\sl S.pombe}} & \multicolumn{11}{c}{{\sl
E.cuniculi}} \\ \hline $q$ & chr.I & chr.II & chr.III & I & II & III & IV & V & VI & VII
& VIII & IX & X & XI\\ \hline 2 & 0.00757 & 0.00743 & 0.00793 & 0.02042 & 0.01953 &
0.01846 & 0.02016 & 0.01968 & 0.02100 & 0.02102 & 0.02057 & 0.01990 & 0.02045 & 0.01953\\
3 & 0.00202 & 0.00200 & 0.00211 & 0.01022 & 0.01027 & 0.01018 & 0.00945 & 0.01067 &
0.01088 & 0.00992 & 0.00982 & 0.00974 & 0.01023 & 0.00989\\ 4 & 0.00204 & 0.00205 &
0.00217 & 0.00565 & 0.00484 & 0.00466 & 0.00448 & 0.00417 & 0.00472 & 0.00469 & 0.00472 &
0.00429 & 0.00445 & 0.00469\\ 5 & 0.00079 & 0.00080 & 0.00102 & 0.00503 & 0.00385 &
0.00425 & 0.00401 & 0.00398 & 0.00406 & 0.00393 & 0.00377 & 0.00349 & 0.00386 & 0.00407\\
6 & 0.00092 & 0.00108 & 0.00121 & 0.00950 & 0.00874 & 0.00831 & 0.00772 & 0.00843 &
0.00807 & 0.00745 & 0.00771 & 0.00669 & 0.00699 & 0.00694\\ 7 & 0.00231 & 0.00219 &
0.00201 & 0.02737 & 0.02726 & 0.02826 & 0.02455 & 0.02626 & 0.02523 & 0.02466 & 0.02305 &
0.02229 & 0.02114 & 0.02128\\ 8 & 0.00428 & 0.00387 & 0.00346 & 0.10920 & 0.10910 &
0.11163 & 0.09981 & 0.10269 & 0.10014 & 0.09642 & 0.09141 & 0.08825 & 0.08316 & 0.08238\\
9 & 0.00955 & 0.00758 & 0.00699 & 0.31844 & 0.32478 & 0.33002 & 0.30823 & 0.30911 &
0.30287 & 0.29743 & 0.29264 & 0.28492 & 0.27174 & 0.27101\\ 10 & 0.03413 & 0.02986 &
0.07854 & 0.43406 & 0.43628 & 0.43629 & 0.44194 & 0.43809 & 0.43332 & 0.43741 & 0.43859 &
0.44005 & 0.43754 & 0.43805\\ 11 & 0.18561 & 0.14578 & 0.37254 & 0.27997 & 0.28007 &
0.27795 & 0.29237 & 0.28930 & 0.29570 & 0.29900 & 0.30620 & 0.31277 & 0.32115 & 0.32204\\
12 & 0.41966 & 0.38547 & 0.39866 & 0.11074 & 0.10991 & 0.10736 & 0.11744 & 0.11686 &
0.12058 & 0.12409 & 0.12828 & 0.13220 & 0.13832 & 0.13953\\ 13 & 0.47562 & 0.41855 &
0.32564 & 0.03657 & 0.03448 & 0.03309 & 0.03791 & 0.03857 & 0.04054 & 0.04062 & 0.04082 &
0.04183 & 0.04613 & 0.04571\\ 14 & 0.42864 & 0.39654 & 0.12658 & 0.01142 & 0.01000 &
0.00975 & 0.01142 & 0.01118 & 0.01200 & 0.01236 & 0.01204 & 0.01244 & 0.01384 & 0.01350\\
15 & 0.17547 & 0.15487 & 0.06246 & 0.00380 & 0.00313 & 0.00269 & 0.00338 & 0.00300 &
0.00376 & 0.00364 & 0.00356 & 0.00344 & 0.00404 & 0.00383\\ 16 & 0.09254 & 0.08754 &
0.02457 & 0.00122 & 0.00109 & 0.00090 & 0.00108 & 0.00101 & 0.00097 & 0.00091 & 0.00122 &
0.00109 & 0.00112 & 0.00126\\ 17 & 0.04369 & 0.03965 & 0.01288 & 0.00045 & 0.00030 &
0.00019 & 0.00032 & 0.00029 & 0.00037 & 0.00035 & 0.00027 & 0.00036 & 0.00043 & 0.00042\\
18 & 0.01485 & 0.00987 & 0.00712 & 0.00023 & 0.00008 & 0.00011 & 0.00008 & 0.00010 &
0.00020 & 0.00012 & 0.00014 & 0.00012 & 0.00013 & 0.00014\\ 19 & 0.00204 & 0.00175 &
0.00121 & 0.00015 & 0.00010 & 0.00005 & 0.00008 & 0.00004 & 0.00005 & 0.00005 & 0.00006 &
0.00003 & 0.00003 & 0.00007\\ 20 & 0.00021 & 0.00018 & 0.00013 & 0.00020 & 0.00005 &
0.00004 & 0.00005 & 0.00001 & 0.00001 & 0.00001 & 0.00005 & 0.00001 & 0.00004 & 0.00002\\
\end{tabular}
\end{ruledtabular}
\end{table*}

Quite often, the expression (\ref{vosst}) is considered to be an evidence of the
Markovian property of an original sequence \cite{durak}, while that is not true. The
expression (\ref{vosst}) is derived with no respect to a structure of the sequence.
Indeed, this expression coincides to the formula for the Markov process of $q$--th order.
The coincidence is not odd; it means, that Markov model of a sequences realizes the
hypothesis of the most probable continuation of a string. We shall discuss this issue
further (see section~\ref{mark}).

\section{Information capacity}\label{inkap}
Information capacity is a measure of deviation of the reconstructed dictionary
(\ref{infcap}) from the real one. The deviation could be measure in various ways. The
approach based on so called ``quality of reconstruction" of dictionary is discussed in
\cite{n1,n2,n3}. A student may implement a regular Euclidean distance to determine the
difference between $\widetilde{W}_{q+1}$ and $W_{q+1}$. Here we explore more sensitive
and more efficient method to detect the difference between the entities, based on the
calculation of mutual entropy.

Mutual entropy of a distribution $\phi$ with respect to a distribution $\psi^{\ast}$ is
defined as
\begin{equation}\label{ue}
\overline{S}\left[\psi^{\ast}|\phi\right] = \sum_{\mu} \phi \cdot \ln
\left(\frac{\phi}{\psi^{\ast}}\right)\; ,
\end{equation}
where $\mu$ is the space of definition of distributions $\phi$ and $\psi^{\ast}$
\cite{bal,obhod}. Here the distribution $\psi^{\ast}$ is the equilibrium one. We shall
define the information capacity in similar way: real frequency $W_q$ should be considered
to be a distribution, while the reconstructed dictionary $\widetilde{W}_q$ should be
considered the ``equilibrium" one. Such definition holds true, since the $q$--support of
the reconstructed dictionary always cntains the $q$--support of the real one.

To determine an information capacity of a frequency dictionary $W_q$, one needs to
develop the relevant reconstructed dictionary $\widetilde{W}_q$ (of the same thickness
$q$). Formula~(\ref{vosst}) is changed for
\begin{subequations}
\label{vs}
\begin{equation}\label{vs:1}
\widetilde{f}_{\nu_1\nu_2\nu_3\ldots \nu_{q-1}\nu_q} = \frac{f_{\nu_1\nu_2\nu_3\ldots
\nu_{q-2}\nu_{q-1}} \times f_{\nu_2\nu_3\ldots \nu_{q-1}\nu_q}}{f_{\nu_2\nu_3\ldots
\nu_{q-2}\nu_{q-2}}}\;,
\end{equation}
\begin{equation}\label{vs:2}
\widetilde{f}_{\nu_1\nu_2} = f_{\nu_1} \times f_{\nu_2}\;,
\end{equation}
\end{subequations}
with (\ref{vs:2}) for the case of $q=2$. The formulae for the reconstruction of a
dictionary of the thickness $q$ over a dictionary of the thickness $s$, $s<q-1$ could be
provided, as well; further we shall keep within the case of the reconstruction of
$\widetilde{W}_q$ over $W_{q-1}$. The expression (\ref{ue}) looks like
\begin{equation}\label{ueslov}
\overline{S}\left[\widetilde{W}_q|W_q\right] = \sum_{\omega} f_{\omega} \cdot \ln
\left(\frac{f_{\omega}}{\widetilde{f}_{\omega}}\right)
\end{equation}
for the case of frequency dictionaries. Substituting (\ref{vs}) into (\ref{ueslov}), one
gets
\begin{subequations}\label{slovs}
\begin{equation}\label{slovs:1}
\overline{S}\left[\widetilde{W}_q|W_q\right]= \sum_{\omega} f_{\omega} \cdot \ln
\left(\frac{f_{\omega}\times f_{\omega''}}{f_{\omega_L}\times f_{\omega_R}}\right)
\end{equation}
\textrm{or}
\begin{equation}\label{slovs:2}
\overline{S}\left[\widetilde{W}_2|W_2\right]= \sum_{\nu_1\nu_2} f_{\nu_1\nu_2} \cdot \ln
\left(\frac{f_{\nu_1\nu_2}}{f_{\nu_1}\times f_{\nu_2}}\right)\;,
\end{equation}
\end{subequations}
here $\omega_L = \nu_1\nu_2\ldots \nu_{q-2}\nu_{q-1}$, $\omega_R = \nu_2\nu_3\ldots
\nu_{q-1}\nu_q$ and $\omega'' = \omega_L \cap \omega_R = \nu_2\nu_3\ldots
\nu_{q-2}\nu_{q-1}$. Expanding the ratio in (\ref{slovs}) into a sum of four terms and
summing up over the ``extra" indices, one gets
\begin{equation}\label{infcap}
\overline{S}_q = 2S_{q-1} - S_q - S_{q-2}\quad \textrm{and}\quad \overline{S}_2 = 2S_1 -
S_2\,.
\end{equation}
The formulae (\ref{infcap}) are changed for $$\overline{S}_q = (q-s+1)S_{s} - S_q -
(q-s)S_{s-1}\quad \textrm{and}\quad \overline{S}_q = qS_1 - S_q$$ for the case of
(\ref{dlinno}).
\begin{table*}
\caption{\label{archea}Information capacity (\ref{infcap}) of nineteen complete genomes
of archea bacteria. $N$ is the length of a genome.}
\begin{ruledtabular}
\begin{tabular}{r|l|r|cccccccc}
\multicolumn{1}{c} {}& \multicolumn{1}{c}{} & \multicolumn{1}{c}{} & \multicolumn{8}{c}
{Frequency thickness} \\ \multicolumn{1}{c}{Entry}  & \multicolumn{1}{c}{Species} &
\multicolumn{1}{c}{$N$} & $q=2$ & $q=3$ & $q=4$ & $q=5$ & $q=6$ & $q=7$ & $q=8$ & $q=9$\\
\hline BA000002 & A.pernix K1 & 1669695 & 0.015290 & 0.007944 & 0.012311 & 0.006464 &
0.005223 & 0.008679 & 0.015346 & 0.053446\\ AE000782 & {\sl A.fulgidus DSM 4304} &
2178400 & 0.018170 & 0.007606 & 0.011233 & 0.008505 & 0.006617 & 0.008051 & 0.013013 &
0.041986\\ AE004437 & {\sl Halobacterium sp}.  & 2014239 & 0.028372 & 0.007114 & 0.013193
& 0.008810 & 0.013955 & 0.013883 & 0.016140 & 0.044798\\ AE000666 & {\sl
M.thermoautotrophicum} $\delta$ H & 1751377 & 0.021919 & 0.012330 & 0.009047 & 0.006213 &
0.007599 & 0.009382 & 0.016337 & 0.052717\\ L77117 & {\sl M.jannaschii DSM 2661} &
1664970 & 0.020938 & 0.007459 & 0.013321 & 0.004747 & 0.006425 & 0.009844 & 0.016392 &
0.046324\\ BX950229 & {\sl M.maripaludis} & 1661137 & 0.015925 & 0.006152 & 0.006783 &
0.005068 & 0.005482 & 0.008590 & 0.017090 & 0.054143\\ AE009439 & {\sl M.kandleri AV19} &
1694969 & 0.016257 & 0.008819 & 0.006385 & 0.003698 & 0.003846 & 0.007801 & 0.014203 &
0.052200\\ AE010299 & {\sl M.acetivorans} & 5751492 & 0.014008 & 0.010528 & 0.003869 &
0.001678 & 0.001791 & 0.004272 & 0.005894 & 0.019365\\ AE008384 & {\sl M.mazei} & 4096345
& 0.015407 & 0.013163 & 0.004834 & 0.001999 & 0.002125 & 0.004579 & 0.006651 & 0.022871\\
AE017199 & {\sl N.equitans} & 490885 & 0.018216 & 0.012787 & 0.008837 & 0.004529 &
0.007170 & 0.016564 & 0.044161 & 0.122603\\ AE017261 & {\sl P.torridus DSM 9790} &
1545895 & 0.015631 & 0.014016 & 0.010894 & 0.008151 & 0.006663 & 0.008954 & 0.017796 &
0.056461\\ AE009441 & P.aerophilum & 2222430 & 0.007622 & 0.014392 & 0.010685 & 0.010080
& 0.004820 & 0.006798 & 0.012019 & 0.040344\\ AL096836 & {\sl P.abyssi} & 1765118 &
0.015921 & 0.008895 & 0.005493 & 0.005236 & 0.004513 & 0.006985 & 0.014085 & 0.049796\\
AE009950 & {\sl P.furiosus DSM 3638} & 1908256 & 0.021224 & 0.005685 & 0.003697 &
0.003117 & 0.003536 & 0.006347 & 0.013273 & 0.046818\\ BA000001 & {\sl P.horikoshii} &
1738505 & 0.019283 & 0.007373 & 0.004917 & 0.003867 & 0.003366 & 0.006636 & 0.013916 &
0.049387\\ AE006641 & {\sl S.solfataricus} & 2992245 & 0.007988 & 0.003279 & 0.004015 &
0.002656 & 0.002072 & 0.005709 & 0.012057 & 0.039543\\ BA000023 & {\sl S.tokodaii}
 & 2694756 & 0.009540 & 0.002887 & 0.004678 & 0.002896 & 0.002598 & 0.005675 &
0.010279 & 0.034593\\ AL139299 & {\sl Th.acidophilum} & 1564906 & 0.010168 & 0.012824 &
0.005046 & 0.003288 & 0.003858 & 0.006748 & 0.014218 & 0.055396\\ BA000011 & {\sl
Th.volcanium} & 1584804 & 0.005425 & 0.006657 & 0.002924 & 0.001421 & 0.002066 & 0.005435
& 0.013266 & 0.054005\\
\end{tabular}
\end{ruledtabular}
\end{table*}

\subsection{Some properties of information capacity}\label{infkap1}
The information capacity defined according to (\ref{infcap}) exhibits some peculiarities
making the capacity a powerful tool for a study of symbol sequences. Let's consider an
estimation for the maximal value of the information capacity (\ref{infcap}). It is
evident, that the maximal level of the information capacity would be observed for the
case where the reconstructed dictionary $\widetilde{W}_q$ has the maximal absolute
entropy, i.e. $\widetilde{f}_{\omega_i} = \widetilde{f}_{\omega_j}, \forall (i, j)$,
while the real frequency of the same thickness must be as far from equilibrium, as
possible.

Consider an infinitely long periodical sequence from two-letter alphabet $\{0,1\}$:
$$\ldots01010101\ldots\; ,$$ with dictionary $W_1$ having two words (these are the
symbols) with equal frequencies. The dictionary $W_2$ has only two words: $01$ and $10$,
with equal frequencies, so that $f_{01} = f_{10} = 1/2$, while $f_{11} = f_{00} = 0$.
Formula (\ref{slovs:2}) (see also (\ref{infcap})) yields $\overline{S}_2 = \ln2$.
Consider, then, an infinite periodical sequence $$\ldots\underbrace{1100}_n\ldots\; .$$
Again, this sequence exhibits an equilibrium frequency dictionary $W_2$ and
quasi-equilibrium dictionary $W_3$; formula (\ref{slovs:1}) yields the same value of
$\overline{S}_3 = \ln2$. Similar sequence could be figured out for any equilibrium
dictionary $W_q$; thus, maximal value of (\ref{slovs}) is equal to $\ln 2$ for any $q$.

The sequence from four-letter alphabet $\aleph = \{\mathsf{A}, \mathsf{C}, \mathsf{G},
\mathsf{T}\}$ with equilibrium dictionary $W_1$, and quasi-equilibrium dictionary $W_2$
is evident: $$\ldots\mathsf{ATGCATGCATGC}\ldots\;.$$ Formula (\ref{slovs:2}) yields
$\overline{S}_2 = 2\ln2$. An infinite periodical sequence $(\mathsf{AACCGGTTGAGCATCT})_n$
provides the same pattern of dictionaries, with $\overline{S}_3 = 2\ln2$. Going this way,
one obtains $\overline{S}_q = 2\ln2$, for any $q$.

If a sequence is arranged from an alphabet $\aleph$ of the cardinality $M$, then the
upper level of information capacity (\ref{slovs}) for such sequence is equal to
$$\overline{S}_q = \frac{M\cdot \ln2}{2}\;.$$ The sense of this relation is clear and
obvious: that is the indeterminacy of a choice of a word of the length $q$ from the
subset of equally distributed ones, with respect to the fact that only a half of all
possible words have positive frequency.

\subsection{Information capacity and redundancy of sequences}\label{dz}
Intuitively, redundancy is a measure of an excess of information content observed within
a sequence. Traditionally, a redundancy is defined through a two-symbol correlations, or
two-symbol entropy calculation, in comparison to a single symbol distribution
\cite{red1,red2,red3,red4,red5}. A study of information capacity of frequency
dictionaries provides a researcher with more advanced definition of the sequence
redundancy.
\begin{table*}
\caption{\label{euk}Information capacity (\ref{infcap}) of several eukaryotic genomes.}
\begin{ruledtabular}
\begin{tabular}{r|c|r|cccccccc}
\multicolumn{1}{c} {}& \multicolumn{1}{c}{} & \multicolumn{1}{c}{} & \multicolumn{8}{c}
{Frequency thickness} \\ \multicolumn{1}{c}{Entry}  & \multicolumn{1}{c}{Chromosome} &
\multicolumn{1}{c}{$N$} & $q=2$ & $q=3$ & $q=4$ & $q=5$ & $q=6$ & $q=7$ & $q=8$ & $q=9$\\
\hline \multicolumn{11}{c} {\sl Encephalitozoon cuniculi} \\ \hline CNS06C8G & I & 209982
& 0.020419 & 0.010215 & 0.005647 & 0.005033 & 0.009497 & 0.027366 & 0.109201 & 0.318442\\
CNS07EG9 & II & 197426 & 0.019526 & 0.010266 & 0.004844 & 0.003853 & 0.008743 & 0.027257
& 0.109101 & 0.324783\\ CNS07EGA & III & 194439 & 0.018463 & 0.010184 & 0.004661 &
0.004254 & 0.008309 & 0.028256 & 0.111627 & 0.330016\\ CNS07EGB & IV & 218329 & 0.020161
& 0.009449 & 0.004475 & 0.004010 & 0.007722 & 0.024553 & 0.099807 & 0.308234\\ CNS07EGC &
V & 251002 & 0.019684 & 0.010673 & 0.004171 & 0.003981 & 0.008430 & 0.026259 & 0.102691 &
0.309111\\ AL590446 & VI & 211018 & 0.020997 & 0.010878 & 0.004720 & 0.004064 & 0.008074
& 0.025227 & 0.100142 & 0.302868\\ AL590447 & VII & 220294 & 0.021018 & 0.009918 &
0.004688 & 0.003929 & 0.007445 & 0.024660 & 0.096422 & 0.297431\\ AL590448 & VIII &
226576 & 0.020570 & 0.009823 & 0.004718 & 0.003770 & 0.007709 & 0.023050 & 0.091410 &
0.292643\\ AL590445 & IX & 238147 & 0.019905 & 0.009738 & 0.004288 & 0.003486 & 0.006685
& 0.022291 & 0.088252 & 0.284922\\ AL590449 & X & 262797 & 0.020453 & 0.010226 & 0.004451
& 0.003865 & 0.006990 & 0.021142 & 0.083162 & 0.271742\\ AL590450 & XI & 267509 &
0.019526 & 0.009886 & 0.004695 & 0.004067 & 0.006937 & 0.021279 & 0.082380 & 0.271006\\
\hline \multicolumn{11}{c} {\sl Eremothecium gossypii} \\ \hline AE016814 & I & 691920 &
0.004391 & 0.002940 & 0.005624 & 0.001475 & 0.002799 & 0.010756 & 0.029318 & 0.127641\\
AE016815 & II & 867694 & 0.004230 & 0.003054 & 0.004974 & 0.001441 & 0.002521 & 0.009130
& 0.023515 & 0.100758\\ AE016816 & III & 907057 & 0.004661 & 0.002835 & 0.006406 &
0.001625 & 0.002485 & 0.010476 & 0.022534 & 0.096182\\ AE016817 & IV & 1466891 & 0.004174
& 0.002903 & 0.005306 & 0.001186 & 0.001789 & 0.007174 & 0.014014 & 0.056525\\ AE016818 &
V & 1519138 & 0.004194 & 0.002792 & 0.005086 & 0.001239 & 0.001775 & 0.006852 & 0.013755
& 0.054365\\ AE016819 & VI & 1812713 & 0.004247 & 0.002820 & 0.005535 & 0.001184 &
0.001736 & 0.006983 & 0.011623 & 0.044951\\ AE016820 & VII & 1476021 & 0.004276 &
0.002914 & 0.005688 & 0.001366 & 0.001792 & 0.007751 & 0.014080 & 0.056334\\ \hline
\multicolumn{11}{c} {\sl Plasmodium falciparum} \\ \hline AL844501 & 1 & 643292 &
0.004537 & 0.021562 & 0.012614 & 0.016715 & 0.014143 & 0.026339 & 0.043922 & 0.094463\\
AE001362 & 2 & 947102 & 0.005162 & 0.023034 & 0.010135 & 0.011752 & 0.008768 & 0.017372 &
0.030138 & 0.069398\\ AL844502 & 3 & 1060087 & 0.005085 & 0.021935 & 0.010529 & 0.013345
& 0.009207 & 0.016876 & 0.027924 & 0.065252\\ AL844503 & 4 & 1204112 & 0.006206 &
0.022380 & 0.011045 & 0.013383 & 0.010601 & 0.019553 & 0.028004 & 0.065541\\ AL844504 & 5
& 1343552 & 0.005692 & 0.024463 & 0.009726 & 0.011810 & 0.007075 & 0.013638 & 0.021777 &
0.051344\\ AL844505 & 6 & 1418244 & 0.005632 & 0.022289 & 0.009903 & 0.012013 & 0.008739
& 0.016125 & 0.024536 & 0.054656\\ AL844506 & 7 & 1351552 & 0.005754 & 0.021366 &
0.010259 & 0.010999 & 0.007295 & 0.014541 & 0.022528 & 0.055176\\ AL844507 & 8 & 1325595
& 0.006079 & 0.024584 & 0.010025 & 0.012517 & 0.007839 & 0.014781 & 0.022590 & 0.054937\\
AL844508 & 9 & 1541723 & 0.005251 & 0.024587 & 0.009218 & 0.012546 & 0.008096 & 0.014320
& 0.020518 & 0.048233\\ AE014185 & 10 & 1694445 & 0.004614 & 0.023440 & 0.011864 &
0.012760 & 0.009820 & 0.017158 & 0.023704 & 0.049696\\ AE014186 & 11 & 2035250 & 0.005433
& 0.024747 & 0.009713 & 0.011386 & 0.006347 & 0.012405 & 0.016579 & 0.038632\\ AE014188 &
12 & 2271916 & 0.005712 & 0.023763 & 0.009716 & 0.011348 & 0.006377 & 0.011493 & 0.014971
& 0.036721\\ AL844509 & 13 & 2732359 & 0.005614 & 0.022843 & 0.008908 & 0.010828 &
0.006489 & 0.011203 & 0.013794 & 0.031419\\ AE014187 & 14 & 3291006 & 0.005750 & 0.024419
& 0.008603 & 0.010351 & 0.005407 & 0.009717 & 0.011047 & 0.025341\\ \hline
\multicolumn{11}{c} {\sl Trypanosoma brucei} \\ \hline AL929608 & 1 & 1056003 & 0.008134
& 0.011493 & 0.011055 & 0.008356 & 0.013433 & 0.026265 & 0.043841 & 0.119718\\ AE017150 &
2 & 1193931 & 0.008328 & 0.009158 & 0.008691 & 0.004800 & 0.005640 & 0.013662 & 0.035133
& 0.116308\\ \hline \multicolumn{11}{c} {\sl Candida glabrata strain CBS138} \\ \hline
CR380947 & A & 485192 & 0.006112 & 0.002636 & 0.003803 & 0.002073 & 0.003990 & 0.013642 &
0.045155 & 0.174482\\ CR380948 & B & 502101 & 0.006527 & 0.002561 & 0.003240 & 0.001568 &
0.003522 & 0.012129 & 0.042753 & 0.163491\\ CR380949 & C & 558804 & 0.006026 & 0.002585 &
0.003555 & 0.001937 & 0.003717 & 0.012608 & 0.041452 & 0.157846\\ CR380950 & D & 651701 &
0.006884 & 0.002475 & 0.002945 & 0.001584 & 0.002988 & 0.009860 & 0.032530 & 0.128877\\
CR380951 & E & 687501 & 0.006841 & 0.002654 & 0.003418 & 0.001701 & 0.003165 & 0.010837 &
0.033211 & 0.126183\\ CR380952 & F & 927101 & 0.006915 & 0.002329 & 0.002912 & 0.001432 &
0.002149 & 0.007396 & 0.022390 & 0.090687\\ CR380953 & G & 992211 & 0.007020 & 0.002484 &
0.003094 & 0.001407 & 0.002323 & 0.007444 & 0.021144 & 0.085899\\ CR380954 & H & 1050361
& 0.006912 & 0.002489 & 0.003002 & 0.001256 & 0.002165 & 0.006884 & 0.020244 & 0.081305\\
CR380955 & I & 1089401 & 0.006330 & 0.002423 & 0.003010 & 0.001320 & 0.002296 & 0.007551
& 0.021251 & 0.081453\\ CR380956 & J & 1192501 & 0.006948 & 0.002566 & 0.003183 &
0.001570 & 0.002191 & 0.007090 & 0.018903 & 0.073283\\ CR380957 & K & 1302002 & 0.006974
& 0.002343 & 0.002822 & 0.001261 & 0.001924 & 0.005637 & 0.016214 & 0.064765\\ CR380958 &
L & 1440588 & 0.006895 & 0.002597 & 0.002882 & 0.001305 & 0.001999 & 0.006217 & 0.015926
& 0.060181\\ CR380959 & M & 1400893 & 0.006922 & 0.002479 & 0.002850 & 0.001349 &
0.001845 & 0.005611 & 0.014860 & 0.060465\\\hline \multicolumn{11}{c} {\sl Yarrowia
lipolytica strain CLIB99} \\ \hline CR382127 & A & 2303261 & 0.007971 & 0.004445 &
0.004349 & 0.002684 & 0.003795 & 0.006816 & 0.012190 & 0.040135\\ CR382128 & B & 3066374
& 0.007967 & 0.004337 & 0.004082 & 0.002549 & 0.003753 & 0.006128 & 0.009337 & 0.028461\\
CR382129 & C & 3272609 & 0.008314 & 0.004102 & 0.004062 & 0.002532 & 0.003487 & 0.005935
& 0.009274 & 0.028189\\ CR382130 & D & 3633272 & 0.008142 & 0.004317 & 0.004421 &
0.002759 & 0.003827 & 0.006058 & 0.008593 & 0.024927\\ CR382131 & E & 4224103 & 0.008102
& 0.004520 & 0.004068 & 0.002529 & 0.003677 & 0.005642 & 0.007702 & 0.021561\\ CR382132 &
F & 4003362 & 0.007716 & 0.004372 & 0.004336 & 0.002662 & 0.003557 & 0.005920 & 0.007903
& 0.022481\\
\end{tabular}
\end{ruledtabular}
\end{table*}

Consider again the chain (\ref{zep}) of dictionaries. A frequency dictionary $W_q$ is the
redundant one, if it guarantees an unambiguous reconstruction of thicker dictionary.
Critical thickness $d^{\ast}$ of the redundant dictionary could be determined
constructively; $d^{\ast}$ is the thickness yielding uniqueness of any word in it. Some
biological issues concerning the determination of $d^{\ast}$ for various segments of
genes are presented and discussed in \cite{g3,g4,g5,g6}. Definition of redundancy trough
the length of the longest common repeat is simple and transparent; meanwhile, it has a
serious disadvantage. The measure of redundancy tends to $N$ (here $N$ is the length of
entire sequence), for rather simple and obviously redundant sequences, such as periodical
ones.

On the contrary, the calculation of information capacity is free of that discrepancy. As
soon, as $\overline{S}_q = 0$ for some $\overline{q}$, then this thickness of dictionary
should be considered as a redundancy measure. Surely, this specific thickness
$\overline{q}$ depends both on a structure of a sequence (whatever one understands for
that), and on its length.

\subsection{Information capacity of some real genetic systems}\label{realgenes}
Here we present some results of the information capacity (\ref{infcap}) determination for
real nucleotide sequences. We studied the complete genomes of bacteria and eukaryotes;
all the entities are deposited at EMBL--bank.

Table~\ref{pombe} shows the results of the information capacity calculation for the {\sl
Schizosaccharomyces pombe} yeast complete genome and for protozoan {\sl Encephalitozoon
cuniculi} complete genome. It is evident, that the pattern of information capacity
(\ref{infcap}) is bell-shaped. Such pattern results from a finiteness of the length of
the original sequence. Indeed, as $q$ approaches the length of 20 nucleotides, the
abundance of a complete support of a dictionary becomes equal to $4^{20}$; this number
exceeds $10^{12}$; such long genomes are not found yet. The huge number of words follows
in a lack of the greater part of them in a dictionary. This exponentially growing
abundance follows in a degeneration of a thicker frequency dictionary: the greatest
majority of the words occur in a single copy (see section~\ref{dz} and
\cite{g3,g4,g5,g6}).

A study of location and value of maximum of information capacity (\ref{infcap}) makes
sense for the sequences of a close length. Besides, the location (i.e. the thickness $q$)
of the maximum and its value are both sensitive to a structure of a sequence. A
degeneracy of a dictionary follows in a shift of the maximum of (\ref{infcap}) to shorter
words. The genomes are rather diverse, from that point of view.

The behaviour of information capacity for $2 \leq q \leq 9$ is of the greatest interest.
It is evident, that the information capacity (\ref{infcap}) varies non-monotonously. An
excess of $\overline{S}_2$ over $\overline{S}_3$ is a well known fact \cite{n1,n2,n3}
with rather clear biological explanation \cite{r1,r3,witten2001}. The occurrence of two
or three minima in the information capacity pattern is a newly established fact. It
should be said, that a multi-minima pattern is rather widely spread among the studied
genomes, while it is not obligatory. The genome of {\sl E.cuniculi} exibits a single
minimum of the information capacity at $q= 5$, for all chromosomes. Table~\ref{archea}
shows the information capacity for $2 \leq q \leq 9$ for nineteen complete genomes of
archeabacteria. Eleven entities have two minima of the information capacity. This table
presents a new phenomenon: three genomes exhibit an inversion in the information capacity
variation at $2 \leq q \leq 4$. These are {\sl Pyrobaculum aerophilum} (identifier
AE009441), {\sl Thermoplasma acidophilum} (identifier AL139299) and {\sl Thermoplasma
volcanium} (identifier BA000011).

The inversion observed in a family of archeabacteria could also be observed in other
genomes, with various taxonomy. Table~\ref{euk} shows the information capacity determined
for $2 \leq q \leq 9$ over various eukaryotic genomes. Finally, Table~\ref{eubak} present
the most abundant data concerning the behavior of information capacity (\ref{infcap}) for
over 150 complete genomes of eubacteria. Few words should be said towards the format of
this Table. Due to space limitations, all nomenclature of bacteria is provided in a
shortened form. Some lines are identified with the same name of a species; it means, that
such entities belong to different strains, or different serovariants. The detailed
information concerning the taxonomy of the genome could be retrieved from EMBL--bank by
an identifier.

\begin{table*}
\caption{\label{eubak}Information capacity (\ref{infcap}) of several eukaryotic genomes.}
\begin{ruledtabular}
\begin{tabular}{l|l|l|cccccccc}
\multicolumn{1}{c}{Entry}  & \multicolumn{1}{c}{$N$} & \multicolumn{1}{c}{Species} &
$q=2$ & $q=3$ & $q=4$ & $q=5$ & $q=6$ & $q=7$ & $q=8$ & $q=9$\\ \hline CR543861 & 3598621
& {\sl Acinetobacter sp.} & 0.013695 & 0.006754 & 0.003180 & 0.003494 & 0.003362 &
0.003889 & 0.007593 & 0.025350\\ AE007870 & 2841581 & {\sl Ag.tumefaciens} & 0.024522 &
0.012524 & 0.007616 & 0.008113 & 0.004989 & 0.007159 & 0.013111 & 0.042823\\ AE008689 &
2074782 & {\sl Ag.tumefaciens} & 0.024518 & 0.012519 & 0.007616 & 0.008108 & 0.004986 &
0.007162 & 0.013104 & 0.042800\\ AE007869 & 2841490 & {\sl Ag.tumefaciens} & 0.025693 &
0.013720 & 0.008115 & 0.008678 & 0.005462 & 0.007398 & 0.011357 & 0.032478\\ AE008688 &
2075560 & {\sl Ag.tumefaciens} & 0.025694 & 0.013721 & 0.008115 & 0.008678 & 0.005461 &
0.007397 & 0.011352 & 0.032470\\ AE000657 & 1551335 & {\sl Aq.aeolicus VF5} & 0.025086 &
0.013949 & 0.010799 & 0.006867 & 0.005535 & 0.009461 & 0.016737 & 0.057694\\ AE017225 &
5227293 & {\sl B.anthracis} & 0.005011 & 0.004804 & 0.006296 & 0.003584 & 0.002930 &
0.004548 & 0.006860 & 0.019529\\ AE016879 & 5228310 & {\sl B.anthracis} & 0.005011 &
0.004803 & 0.006296 & 0.003586 & 0.002928 & 0.004543 & 0.006852 & 0.019510\\ AE017334 &
5228663 & {\sl B.anthracis} & 0.005013 & 0.004802 & 0.006294 & 0.003584 & 0.002926 &
0.004542 & 0.006852 & 0.019509\\ AE017194 & 5224283 & {\sl B.cereus} & 0.005092 &
0.004713 & 0.006247 & 0.003507 & 0.002941 & 0.004795 & 0.007201 & 0.020537\\ AE016877 &
5411809 & {\sl B.cereus} & 0.004948 & 0.004500 & 0.006028 & 0.003295 & 0.002789 &
0.004601 & 0.007054 & 0.020032\\ BA000004 & 4202352 & {\sl B.halodurans} & 0.011038 &
0.003900 & 0.005231 & 0.002782 & 0.001808 & 0.003647 & 0.007395 & 0.024234\\ AL009126 &
4214630 & {\sl B.subtilis} & 0.016737 & 0.009942 & 0.004373 & 0.001952 & 0.001894 &
0.003744 & 0.006976 & 0.023948\\ AE017355 & 5237682 & {\sl B.thuringiensis} & 0.004957 &
0.004765 & 0.006262 & 0.003542 & 0.002932 & 0.004808 & 0.007532 & 0.021289\\ AE015928 &
6260361 & {\sl B.thetaiotaomicron} & 0.004681 & 0.010665 & 0.006649 & 0.002838 & 0.002634
& 0.003228 & 0.004825 & 0.015111\\ BX897699 & 1931047 & {\sl B.henselae} & 0.015746 &
0.005150 & 0.004615 & 0.002532 & 0.002570 & 0.005103 & 0.013421 & 0.049652\\ BX897700 &
1581384 & {\sl B.quintana} & 0.016423 & 0.004543 & 0.004067 & 0.002225 & 0.002388 &
0.005020 & 0.014100 & 0.056008\\ BX842601 & 3782950 & {\sl B.bacteriovorus } & 0.021078 &
0.010726 & 0.003837 & 0.004967 & 0.004054 & 0.005209 & 0.008686 & 0.024182\\ AE014295 &
2256646 & {\sl B.longum} & 0.013680 & 0.010854 & 0.005459 & 0.007544 & 0.005795 &
0.009393 & 0.013759 & 0.043138\\ BX470250 & 5339179 & {\sl B.bronchiseptica} & 0.019051 &
0.018226 & 0.014946 & 0.009624 & 0.007019 & 0.014473 & 0.010739 & 0.020758\\ BX470249 &
4773551 & {\sl B.parapertussis} & 0.019309 & 0.018475 & 0.015118 & 0.009716 & 0.007072 &
0.014730 & 0.011562 & 0.023433\\ BX470248 & 4086189 & {\sl B.pertussis} & 0.018533 &
0.018289 & 0.014508 & 0.010287 & 0.010869 & 0.026770 & 0.044285 & 0.097603\\ AE000783 &
910724 & {\sl B.burgdorferi} & 0.023227 & 0.002133 & 0.003246 & 0.002873 & 0.003887 &
0.009814 & 0.024221 & 0.076756\\ BA000040 & 9105828 & {\sl B.japonicum} & 0.027282 &
0.009713 & 0.008155 & 0.006690 & 0.004230 & 0.007997 & 0.006765 & 0.013241\\ AE008917 &
2117144 & {\sl B.melitensis, chr.I} & 0.028077 & 0.011210 & 0.006071 & 0.007949 &
0.007784 & 0.007177 & 0.013674 & 0.043426\\ AE008918 & 1177787 & {\sl B.melitensis,
chr.II} & 0.028054 & 0.010904 & 0.006750 & 0.008201 & 0.008306 & 0.008633 & 0.020434 &
0.070601\\ AE014291 & 2107793 & {\sl B.suis 1330, chr.I} & 0.028299 & 0.011342 & 0.006123
& 0.008063 & 0.007878 & 0.007275 & 0.013840 & 0.043651\\ AE014292 & 1207381 & {\sl
B.suis, chr.II} & 0.027777 & 0.010778 & 0.006642 & 0.008118 & 0.008243 & 0.008567 &
0.019980 & 0.069020\\ BA000003 & 640681 & {\sl B.aphidicola} & 0.007608 & 0.004984 &
0.007715 & 0.002464 & 0.004401 & 0.011496 & 0.033328 & 0.100963\\ AE016826 & 615980 &
{\sl B.aphidicola} & 0.004517 & 0.004223 & 0.005281 & 0.002105 & 0.004130 & 0.011172 &
0.034193 & 0.097610\\ AE013218 & 641454 & {\sl B.aphidicola} & 0.010056 & 0.005607 &
0.007001 & 0.002654 & 0.004619 & 0.011923 & 0.032871 & 0.093878\\ AL111168 & 1641481 &
{\sl C.jejuni} & 0.025925 & 0.007195 & 0.008380 & 0.006480 & 0.006136 & 0.009552 &
0.017044 & 0.049726\\ BX248583 & 705557 & {\sl B.floridanus} & 0.004191 & 0.004090 &
0.004874 & 0.002726 & 0.004370 & 0.010925 & 0.030405 & 0.093783\\ AE005673 & 4016947 &
{\sl C.crescentus} & 0.015292 & 0.022869 & 0.012056 & 0.012406 & 0.011076 & 0.028279 &
0.070484 & 0.163516\\ AE002160 & 1072950 & {\sl Ch.muridarum} & 0.015077 & 0.005375 &
0.002794 & 0.002095 & 0.002951 & 0.006255 & 0.019863 & 0.078997\\ AE001273 & 1042519 &
{\sl Ch.trachomatis} & 0.013610 & 0.005905 & 0.003162 & 0.002003 & 0.002976 & 0.006651 &
0.020558 & 0.082006\\ AE015925 & 1173390 & {\sl Ch.caviae} & 0.010397 & 0.005354 &
0.003291 & 0.003658 & 0.004232 & 0.006445 & 0.018200 & 0.072489\\ AE002161 & 1229853 &
{\sl Ch.pneumoniae} & 0.013529 & 0.004222 & 0.002565 & 0.003561 & 0.003454 & 0.006385 &
0.017356 & 0.069847\\ AE001363 & 1230230 & {\sl Ch.pneumoniae} & 0.013530 & 0.004219 &
0.002566 & 0.003566 & 0.003454 & 0.006398 & 0.017363 & 0.069779\\ BA000008 & 1226565 &
{\sl Ch.pneumoniae} & 0.013539 & 0.004219 & 0.002549 & 0.003553 & 0.003438 & 0.006368 &
0.017319 & 0.069720\\ AE009440 & 1225935 & {\sl Ch.pneumoniae} & 0.013506 & 0.004219 &
0.002566 & 0.003564 & 0.003453 & 0.006393 & 0.017342 & 0.069793\\ AE006470 & 2154946 &
{\sl C.tepidum} & 0.020061 & 0.011061 & 0.008408 & 0.006773 & 0.008106 & 0.008369 &
0.012377 & 0.041511\\ AE016825 & 4751080 & {\sl C.violaceum} & 0.019041 & 0.018825 &
0.014015 & 0.009803 & 0.007443 & 0.016733 & 0.011386 & 0.024193\\ AE001437 & 3940880 &
{\sl C.acetobutylicum} & 0.010606 & 0.005123 & 0.004222 & 0.001690 & 0.002404 & 0.005410
& 0.008253 & 0.025033\\ BA000016 & 3031430 & {\sl C.perfringens} & 0.017457 & 0.004479 &
0.005310 & 0.002498 & 0.004307 & 0.008116 & 0.011921 & 0.030875\\ AE015927 & 2799251 &
{\sl C.tetani} & 0.014410 & 0.006102 & 0.006276 & 0.002459 & 0.003594 & 0.006788 &
0.010479 & 0.031232\\ BX248353 & 2488635 & {\sl C.diphtheriae} & 0.007448 & 0.003906 &
0.005199 & 0.004450 & 0.003300 & 0.005425 & 0.010729 & 0.035929\\ BA000035 & 3147090 &
{\sl C.efficiens} & 0.013561 & 0.010039 & 0.011470 & 0.009562 & 0.006170 & 0.009947 &
0.013196 & 0.034160\\ AX114121 & 3309400 & {\sl C.glutamicum} & 0.011670 & 0.005214 &
0.004643 & 0.005185 & 0.003740 & 0.005143 & 0.009104 & 0.027874\\ BA000036 & 3309401 &
{\sl C.glutamicum} & 0.011670 & 0.005214 & 0.004643 & 0.005185 & 0.003740 & 0.005143 &
0.009104 & 0.027873\\ BX927147 & 3282708 & {\sl C.glutamicum} & 0.011725 & 0.005243 &
0.004649 & 0.005196 & 0.003758 & 0.005142 & 0.009162 & 0.028107\\ AE016828 & 1995275 &
{\sl C.burnetii strain} & 0.020865 & 0.001531 & 0.004846 & 0.001590 & 0.002105 & 0.004843
& 0.012999 & 0.047250\\ AE000513 & 2648638 & {\sl D.radiodurans, chr.1} & 0.010846 &
0.008048 & 0.013872 & 0.009601 & 0.016747 & 0.013908 & 0.015656 & 0.038380\\ AE001825 &
412348 & {\sl D.radiodurans, chr.2} & 0.010821 & 0.007962 & 0.014203 & 0.009626 &
0.018188 & 0.022831 & 0.054850 & 0.149147\\ AE017285 & 3570858 & {\sl D.vulgaris} &
0.010306 & 0.013164 & 0.009729 & 0.010660 & 0.007107 & 0.007437 & 0.009646 & 0.026623\\
AE016830 & 3218031 & {\sl E.faecalis} & 0.013746 & 0.002634 & 0.005872 & 0.002905 &
0.003065 & 0.004865 & 0.008791 & 0.028356\\ BX950851 & 5231428 & {\sl E.carotovora} &
0.011198 & 0.010238 & 0.006360 & 0.005196 & 0.003841 & 0.003902 & 0.006142 & 0.018491\\
AE014075 & 4639675 & {\sl E.coli} & 0.011808 & 0.012713 & 0.008453 & 0.004938 & 0.004337
& 0.003890 & 0.006233 & 0.018843\\ U00096 & 5498450 & {\sl E.coli K-12} & 0.012761 &
0.012604 & 0.008689 & 0.005368 & 0.004935 & 0.004300 & 0.007042 & 0.021171\\ AE005174 &
5528970 & {\sl E.coli} & 0.011766 & 0.012788 & 0.008379 & 0.004734 & 0.004253 & 0.004040
& 0.006233 & 0.018996\\ BA000007 & 2174500 & {\sl E.coli} & 0.011825 & 0.012822 &
0.008422 & 0.004794 & 0.004350 & 0.004063 & 0.006187 & 0.018753\\
\end{tabular}
\end{ruledtabular}
\end{table*}
\begin{table*}
\begin{ruledtabular}
\begin{tabular}{l|l|l|cccccccc}
AE009951 & 3814139 & {\sl F.nucleatum} & 0.021164 & 0.005779 & 0.009750 & 0.003532 &
0.006993 & 0.009325 & 0.013460 & 0.034533\\ AE017180 & 4659019 & {\sl G.sulfurreducens} &
0.009219 & 0.007008 & 0.010504 & 0.005446 & 0.004269 & 0.008725 & 0.009014 &
0.025514\\BA000045 & 1698955 & {\sl G.violaceus} & 0.010012 & 0.009740 & 0.010249 &
0.006336 & 0.006289 & 0.008805 & 0.009177 & 0.021508\\ AE017143 & 1830138 & {\sl
H.ducreyi} & 0.014763 & 0.004629 & 0.006226 & 0.004448 & 0.004560 & 0.006575 & 0.015308 &
0.055473\\ L42023 & 1799146 & {\sl H.influenzae} & 0.017150 & 0.004225 & 0.007368 &
0.005328 & 0.006523 & 0.007372 & 0.015226 & 0.050344\\ AE017125 & 1667867 & {\sl
H.hepaticus} & 0.018702 & 0.014497 & 0.014539 & 0.008144 & 0.008877 & 0.007023 & 0.013430
& 0.041564\\ AE000511 & 1643831 & {\sl H.pylori} & 0.030840 & 0.015429 & 0.015576 &
0.012382 & 0.008501 & 0.008442 & 0.016689 & 0.050142\\ AE001439 & 1992676 & {\sl
H.pylori} & 0.030502 & 0.015549 & 0.015428 & 0.012521 & 0.008944 & 0.008466 & 0.017338 &
0.051273\\ AE017198 & 3308274 & {\sl L.johnsonii} & 0.009143 & 0.004275 & 0.004558 &
0.002686 & 0.003695 & 0.006929 & 0.014344 & 0.047679\\ AL935263 & 2365589 & {\sl
L.plantarum} & 0.008293 & 0.010481 & 0.003223 & 0.001675 & 0.002355 & 0.003553 & 0.007938
& 0.026947\\ AE005176 & 350181 & {\sl L.lactis} & 0.016683 & 0.003346 & 0.006059 &
0.002677 & 0.004539 & 0.006436 & 0.013119 & 0.042277\\ AE010300 & 4277185 & {\sl
L.interrogans, chr.I} & 0.023489 & 0.005030 & 0.004597 & 0.003392 & 0.003288 & 0.006140 &
0.010908 & 0.028291\\ AE010301 & 4332241 & {\sl L.interrogans, chr.II} & 0.023718 &
0.005441 & 0.004555 & 0.004094 & 0.006199 & 0.017972 & 0.063814 & 0.187752\\ AE016824 &
358943 & {\sl L.interrogans, chr.II} & 0.024007 & 0.005584 & 0.004553 & 0.004111 &
0.006322 & 0.017950 & 0.064234 & 0.188090\\ AE016823 & 3011208 & {\sl L.interrogans} &
0.023558 & 0.005155 & 0.004588 & 0.003391 & 0.003247 & 0.005877 & 0.010228 & 0.026687\\
AL592022 & 2944528 & {\sl L.innocua} & 0.011179 & 0.003573 & 0.005686 & 0.003151 &
0.003294 & 0.005174 & 0.009607 & 0.030576\\ AL591824 & 2905310 & {\sl L.monocytogenes} &
0.011396 & 0.003882 & 0.005642 & 0.003037 & 0.003048 & 0.005090 & 0.009756 & 0.031100\\
AE017262 & 7036071 & {\sl L.monocytogenes} & 0.011505 & 0.003916 & 0.005730 & 0.003086 &
0.003073 & 0.004978 & 0.009834 & 0.031566\\ BA000012 & 4829781 & {\sl M.loti} & 0.024233
& 0.015134 & 0.008290 & 0.007213 & 0.004327 & 0.006993 & 0.006814 & 0.015281\\ AE016958 &
4345492 & {\sl M.avium} & 0.014324 & 0.015800 & 0.014565 & 0.009668 & 0.006895 & 0.012516
& 0.009742 & 0.021269\\ BX248333 & 3268203 & {\sl M.bovis} & 0.011931 & 0.013305 &
0.006231 & 0.005685 & 0.003712 & 0.005707 & 0.008096 & 0.021289\\ AL450380 & 4403836 &
{\sl M.leprae} & 0.006510 & 0.009074 & 0.003134 & 0.002974 & 0.001870 & 0.003343 &
0.007997 & 0.026873\\ AE000516 & 4411532 & {\sl M.tuberculosis} & 0.011986 & 0.013288 &
0.006198 & 0.005666 & 0.003707 & 0.005629 & 0.007962 & 0.020983\\ AL123456 & 996422 &
{\sl M.tuberculosis} & 0.011921 & 0.013306 & 0.006187 & 0.005689 & 0.003735 & 0.005692 &
0.008128 & 0.021367\\ AE015450 & 580074 & {\sl M.gallisepticum} & 0.006701 & 0.015753 &
0.006583 & 0.004337 & 0.004985 & 0.010510 & 0.027155 & 0.079118\\ L43967 & 777079 & {\sl
M.genitalium} & 0.022383 & 0.011217 & 0.006594 & 0.004153 & 0.006728 & 0.014388 &
0.039278 & 0.106222\\ AE017308 & 1211703 & {\sl M.mobile} & 0.025882 & 0.002351 &
0.006275 & 0.004904 & 0.006123 & 0.012948 & 0.028369 & 0.071083\\ BX293980 & 1358633 &
{\sl M.mycoides} & 0.013469 & 0.005017 & 0.007949 & 0.005886 & 0.008636 & 0.018970 &
0.036735 & 0.074542\\ BA000026 & 816394 & {\sl M.penetrans} & 0.019534 & 0.004346 &
0.005586 & 0.004174 & 0.005411 & 0.009853 & 0.017731 & 0.044553\\ U00089 & 963879 & {\sl
M.pneumoniae} & 0.018826 & 0.007344 & 0.005417 & 0.004906 & 0.006282 & 0.013033 &
0.035947 & 0.115842\\ AL445566 & 2272351 & {\sl M.pulmonis} & 0.031785 & 0.003381 &
0.006229 & 0.004446 & 0.006995 & 0.012152 & 0.025206 & 0.062292\\ AL157959 & 2184406 &
{\sl N.meningitidis} & 0.029480 & 0.010012 & 0.014966 & 0.012269 & 0.012474 & 0.016027 &
0.022912 & 0.054354\\ AE002098 & 2812094 & {\sl N.meningitidis} & 0.028833 & 0.010041 &
0.014901 & 0.011777 & 0.011893 & 0.015318 & 0.021837 & 0.053141\\ AL954747 & 6413771 &
{\sl N.europaea} & 0.012756 & 0.015878 & 0.005756 & 0.003580 & 0.002610 & 0.004525 &
0.010362 & 0.035089\\ BA000019 & 3630528 & {\sl Nostoc sp.} & 0.007019 & 0.004369 &
0.007128 & 0.005647 & 0.008247 & 0.004818 & 0.006080 & 0.015953\\ BA000028 & 860631 &
{\sl O.iheyensis} & 0.004552 & 0.003903 & 0.004959 & 0.002415 & 0.001995 & 0.004403 &
0.008011 & 0.026023\\ AP006628 & 2414465 & {\sl Phytoplasma OY-M} & 0.024347 & 0.002811 &
0.009253 & 0.004726 & 0.006011 & 0.013248 & 0.034387 & 0.093834\\ BX908798 & 2257487 &
{\sl P.symbiont UWE25} & 0.019051 & 0.000800 & 0.002495 & 0.001469 & 0.002191 & 0.005323
& 0.011907 & 0.040491\\ AE004439 & 5064019 & {\sl P.multocida} & 0.013903 & 0.003047 &
0.005737 & 0.004126 & 0.004863 & 0.005956 & 0.011835 & 0.040879\\ CR354531 & 4085304 &
{\sl Ph.profundum, chr.1} & 0.007148 & 0.006235 & 0.004428 & 0.002712 & 0.002584 &
0.004022 & 0.008711 & 0.028081\\ CR354532 & 2237943 & {\sl Ph.profundum, chr.2} &
0.007086 & 0.004927 & 0.004542 & 0.002733 & 0.002916 & 0.005705 & 0.015656 & 0.051976\\
BX470251 & 5688987 & {\sl Ph.luminescens} & 0.008510 & 0.006521 & 0.005386 & 0.002895 &
0.003332 & 0.004364 & 0.008499 & 0.022554\\ BX119912 & 7145576 & {\sl Pirellula sp.} &
0.017458 & 0.001464 & 0.003464 & 0.001879 & 0.003685 & 0.005981 & 0.013366 & 0.048556\\
AE015924 & 2343476 & {\sl P.gingivalis} & 0.007081 & 0.010317 & 0.006542 & 0.004243 &
0.003609 & 0.006069 & 0.013509 & 0.044284\\ BX548174 & 2410873 & {\sl P.marinus} &
0.018590 & 0.005901 & 0.003325 & 0.002806 & 0.002921 & 0.004667 & 0.009789 & 0.034915\\
BX548175 & 1751080 & {\sl P.marinus} & 0.014293 & 0.013618 & 0.008557 & 0.005447 &
0.004080 & 0.004461 & 0.006502 & 0.019154\\ AE017126 & 1657990 & {\sl P.marinus} &
0.017482 & 0.001875 & 0.002785 & 0.002210 & 0.003777 & 0.005237 & 0.012796 & 0.047272\\
AE004091 & 6264403 & {\sl P.aeruginosa} & 0.011202 & 0.021308 & 0.014929 & 0.012606 &
0.007620 & 0.013218 & 0.010158 & 0.019504\\ AE015451 & 6181863 & {\sl P.putida} &
0.012929 & 0.016990 & 0.007108 & 0.008053 & 0.004959 & 0.007802 & 0.008244 & 0.018798\\
AE016853 & 6397126 & {\sl P.syringae} & 0.015843 & 0.012798 & 0.005311 & 0.006846 &
0.005050 & 0.005531 & 0.007981 & 0.019878\\ AL646052 & 3716413 & {\sl R.solanacearum} &
0.021411 & 0.013736 & 0.010764 & 0.008860 & 0.005842 & 0.012073 & 0.011065 & 0.027009\\
AE006914 & 5459213 & {\sl R.conorii} & 0.007552 & 0.005067 & 0.006533 & 0.003395 &
0.004597 & 0.009335 & 0.021696 & 0.068644\\ AJ235269 & 1268755 & {\sl R.prowazekii} &
0.005170 & 0.002927 & 0.004268 & 0.002387 & 0.003076 & 0.007000 & 0.019457 & 0.067786\\
AL513382 & 1111523 & {\sl S.enterica} & 0.013751 & 0.013405 & 0.008223 & 0.005347 &
0.003932 & 0.004370 & 0.006655 & 0.019728\\ AE014613 & 4791961 & {\sl S.enterica} &
0.013820 & 0.013388 & 0.008231 & 0.005328 & 0.003932 & 0.004410 & 0.006684 & 0.019829\\
AE006468 & 4809037 & {\sl S.typhimurium} & 0.014293 & 0.013618 & 0.008557 & 0.005447 &
0.004080 & 0.004461 & 0.006502 & 0.019154\\ AE014299 & 4857432 & {\sl Sh.oneidensis} &
0.010458 & 0.006262 & 0.004642 & 0.004382 & 0.003284 & 0.003841 & 0.007949 & 0.023929\\
AE014073 & 4969803 & {\sl Sh.flexneri} & 0.011760 & 0.012701 & 0.008109 & 0.005166 &
0.005014 & 0.005425 & 0.011490 & 0.031813\\ AE005674 & 4607203 & {\sl Sh.flexneri} &
0.011748 & 0.012673 & 0.008080 & 0.005184 & 0.005035 & 0.005501 & 0.011795 & 0.032345\\
AL591688 & 4599354 & {\sl S.meliloti} & 0.026681 & 0.010958 & 0.007164 & 0.006438 &
0.003565 & 0.007529 & 0.009891 & 0.027683\\ BX571856 & 2902619 & {\sl St.aureus} &
0.005913 & 0.002068 & 0.005134 & 0.002628 & 0.003085 & 0.005770 & 0.010129 & 0.032952\\
BX571857 & 2799802 & {\sl St.aureus} & 0.005964 & 0.002117 & 0.005290 & 0.002717 &
0.003144 & 0.005981 & 0.010413 & 0.033942\\
\end{tabular}
\end{ruledtabular}
\end{table*}
\begin{table*}
\begin{ruledtabular}
\begin{tabular}{l|l|l|cccccccc}
BA000017 & 2820462 & {\sl St.aureus} & 0.005854 & 0.002085 & 0.005151 & 0.002687 &
0.003078 & 0.005765 & 0.010071 & 0.032912\\ BA000033 & 2878529 & {\sl St.aureus} &
0.005982 & 0.002104 & 0.005273 & 0.002719 & 0.003128 & 0.005970 & 0.010400 & 0.033816\\
BA000018 & 2814816 & {\sl St.aureus} & 0.005914 & 0.002130 & 0.005207 & 0.002716 &
0.003127 & 0.005906 & 0.010446 & 0.034171\\ AE015929 & 2499279 & {\sl St.epidermidis} &
0.003767 & 0.001716 & 0.004630 & 0.002240 & 0.002595 & 0.006244 & 0.011690 & 0.038364\\
AE009948 & 2160267 & {\sl St.agalactiae} & 0.006904 & 0.002936 & 0.005024 & 0.001858 &
0.002747 & 0.005824 & 0.013064 & 0.044783\\ AL732656 & 2211485 & {\sl St.agalactiae} &
0.007217 & 0.002855 & 0.005103 & 0.001912 & 0.002754 & 0.005794 & 0.012886 & 0.044199\\
AE014133 & 2030921 & {\sl St.mutans} & 0.013456 & 0.004012 & 0.005658 & 0.002547 &
0.002924 & 0.005414 & 0.012257 & 0.044677\\ AE007317 & 2038615 & {\sl St.pneumoniae} &
0.011668 & 0.004819 & 0.005481 & 0.002696 & 0.003326 & 0.006172 & 0.013436 & 0.046620\\
AE005672 & 2160837 & {\sl St.pneumoniae} & 0.011575 & 0.004695 & 0.005364 & 0.002744 &
0.003515 & 0.007163 & 0.014714 & 0.047388\\ AE004092 & 1852441 & {\sl St.pyogenes} &
0.010333 & 0.004626 & 0.004952 & 0.002287 & 0.002627 & 0.005484 & 0.013313 & 0.048073\\
AE014074 & 1900521 & {\sl St.pyogenes} & 0.010106 & 0.004583 & 0.004939 & 0.002203 &
0.002613 & 0.005468 & 0.013328 & 0.048020\\ BA000034 & 1895017 & {\sl St.pyogenes} &
0.010166 & 0.004620 & 0.004978 & 0.002213 & 0.002580 & 0.005414 & 0.013187 & 0.047412\\
AE009949 & 1894275 & {\sl St.pyogenes} & 0.010180 & 0.004543 & 0.004911 & 0.002241 &
0.002626 & 0.005417 & 0.013576 & 0.048444\\ BA000030 & 9025608 & {\sl St.avermitilis} &
0.011393 & 0.010642 & 0.010511 & 0.007005 & 0.006183 & 0.009745 & 0.006162 & 0.012784\\
AL645882 & 8667507 & {\sl St.coelicolor} & 0.011072 & 0.011860 & 0.012735 & 0.007970 &
0.007255 & 0.011645 & 0.006942 & 0.013581\\ BX548020 & 2434428 & {\sl Synechococcus sp.}
& 0.017464 & 0.011350 & 0.007426 & 0.005454 & 0.003216 & 0.007177 & 0.011822 & 0.038270\\
BA000022 & 3573470 & {\sl Synechocystis sp.} & 0.023278 & 0.004463 & 0.008626 & 0.009390
& 0.006909 & 0.006308 & 0.009156 & 0.026581\\ AE008691 & 2689445 & {\sl Th.tengcongensis}
& 0.018306 & 0.007426 & 0.006789 & 0.002851 & 0.003094 & 0.006214 & 0.011581 & 0.038095\\
BA000039 & 2593857 & {\sl Th.elongatus} & 0.015134 & 0.004435 & 0.008833 & 0.008839 &
0.007238 & 0.007249 & 0.013275 & 0.040696\\ AE000512 & 1860725 & {\sl Th.maritima} &
0.028543 & 0.017235 & 0.006157 & 0.003558 & 0.003491 & 0.006186 & 0.013606 & 0.047372\\
AE017221 & 1894877 & {\sl Th.thermophilus} & 0.030010 & 0.028048 & 0.020868 & 0.021224 &
0.014340 & 0.019560 & 0.020008 & 0.040026\\ AE017226 & 2843201 & {\sl T.denticola} &
0.020824 & 0.012443 & 0.007493 & 0.003559 & 0.002909 & 0.005231 & 0.009348 & 0.031935\\
AE000520 & 1138011 & {\sl T.pallidum} & 0.008565 & 0.013489 & 0.004322 & 0.002325 &
0.002653 & 0.006578 & 0.018883 & 0.077105\\ AE014184 & 925938 & {\sl T.whipplei} &
0.005901 & 0.006857 & 0.003644 & 0.002028 & 0.004167 & 0.008065 & 0.026251 & 0.103984\\
BX072543 & 927303 & {\sl T.whipplei} & 0.005888 & 0.006799 & 0.003663 & 0.002031 &
0.004196 & 0.008171 & 0.026376 & 0.103937\\ AF222894 & 751719 & {\sl U.urealyticum} &
0.012392 & 0.005304 & 0.008392 & 0.003739 & 0.005607 & 0.013331 & 0.030211 & 0.081295\\
AE003852 & 2961149 & {\sl V.cholerae, chr.I } & 0.012525 & 0.008069 & 0.003458 & 0.005102
& 0.003751 & 0.004211 & 0.009380 & 0.031341\\ AE003853 & 1072315 & {\sl V.cholerae,
chr.II} & 0.013503 & 0.006883 & 0.003994 & 0.005358 & 0.005951 & 0.010563 & 0.026775 &
0.088844\\ BA000031 & 3288558 & {\sl V.parahaemolyticus, chr.1} & 0.010957 & 0.006585 &
0.003386 & 0.003249 & 0.002822 & 0.003810 & 0.008719 & 0.028953\\ BA000032 & 1877212 &
{\sl V.parahaemolyticus, chr.2} & 0.012741 & 0.005847 & 0.003816 & 0.003567 & 0.003113 &
0.004436 & 0.012061 & 0.045329\\ AE016795 & 3281945 & {\sl V.vulnificus, chr.I} &
0.012698 & 0.006581 & 0.003817 & 0.003317 & 0.002780 & 0.003857 & 0.009136 & 0.029434\\
AE016796 & 1844853 & {\sl V.vulnificus, chr.II} & 0.014850 & 0.006728 & 0.004751 &
0.004073 & 0.003228 & 0.004387 & 0.012116 & 0.045682\\ BA000037 & 3354505 & {\sl
V.vulnificus. chr.I} & 0.012488 & 0.006641 & 0.003806 & 0.003326 & 0.002752 & 0.003844 &
0.009066 & 0.029050\\ BA000038 & 1857073 & {\sl V.vulnificus. chr.II} & 0.014921 &
0.006935 & 0.004853 & 0.004106 & 0.003306 & 0.004377 & 0.012018 & 0.045477\\ BA000021 &
697724 & {\sl W.glossinidia} & 0.014766 & 0.007419 & 0.006667 & 0.002547 & 0.005068 &
0.011597 & 0.029992 & 0.080185\\ AE017196 & 1267782 & {\sl W.endosymbiont} & 0.009537 &
0.003220 & 0.003734 & 0.002173 & 0.004236 & 0.009704 & 0.024983 & 0.077065\\ BX571656 &
2110355 & {\sl W.succinogenes} & 0.028493 & 0.021856 & 0.012718 & 0.008349 & 0.006376 &
0.006379 & 0.013819 & 0.043435\\ AE008923 & 5175554 & {\sl X.axonopodis} & 0.024169 &
0.014838 & 0.007583 & 0.008203 & 0.004462 & 0.008442 & 0.008844 & 0.020333\\ AE008922 &
5076188 & {\sl X.campestris} & 0.023799 & 0.014241 & 0.008130 & 0.008585 & 0.004921 &
0.009403 & 0.009628 & 0.022067\\ AE003849 & 2679306 & {\sl X.fastidiosa} & 0.011826 &
0.003955 & 0.004727 & 0.003342 & 0.003024 & 0.004914 & 0.009535 & 0.033301\\ AE009442 &
2519802 & {\sl X.fastidiosa} & 0.011900 & 0.004018 & 0.004997 & 0.003470 & 0.003060 &
0.005249 & 0.010488 & 0.036900\\ AE017042 & 4595065 & {\sl Y.pestis} & 0.009359 &
0.008596 & 0.005810 & 0.003897 & 0.003320 & 0.003660 & 0.007726 & 0.024715\\ AE009952 &
4600755 & {\sl Y.pestis} & 0.009320 & 0.008555 & 0.005798 & 0.003861 & 0.003332 &
0.003717 & 0.008030 & 0.025373\\ AL590842 & 4653728 & {\sl Y.pestis} & 0.009318 &
0.008568 & 0.005805 & 0.003873 & 0.003394 & 0.003911 & 0.008721 & 0.026808\\
\end{tabular}
\end{ruledtabular}
\end{table*}

\section{Discussion}\label{disk}
A researcher capitalizes a lot from the studies of the statistical properties of
nucleotide sequences. Here we propose a novel approach towards the definition of the
information capacity of a frequency dictionary of a sequence. The key idea of the
information capacity definition is the comparison of real and expected frequency of
considerably short strings occurred within a symbol sequence. A definition of an expected
frequency is the basic problem in the studies of information properties of such entities.

Basically, there are two approaches to identify an expected frequency. The former is to
change a sequence under consideration for some surrogate entity with known (or specially
prepared) statistical properties, say, consider a realization of some random process
\cite{durak,borhy}. The latter is to figure out the most expected continuation of a
string keeping within the information available at the frequency dictionary, only.
Changing an original sequence for surrogate one, a student involves into a study outer,
additional information. Such intrusion of the additional information may conflict with
reliability of the retrieved knowledge and conspire some fine properties of the original
sequence.

Studying the statistical properties of symbol sequences, researchers quite often restrict
themselves with the consideration of mono- and dinucleotide distribution
\cite{z3,z4,z5,z6,a1,a2,a3,a4}. A breakthrough in that direction results from the
fundamental studies in Boltzmann's equation \cite{kg1,kg2,kg3,kg4,kg5,kg6}, which were
successfully converted into the field of bioinformatics \cite{n1,n2,n3,kitai}. In this
paper, we implemented the version of the method of invariant manifolds for figuring out
the formula for the most expected continuation of a string. Since the strings are
discrete objects, the formula~(\ref{infcap}) becomes the exact solution, on the contrary
to the situation of a typical physical situation \cite{kirk,kg2,kg4}.

Zero information capacity of a frequency dictionary $W_q$ means the exact and unambiguous
extension of the given dictionary into any thicker one $W_k$, $k > q$. This point
provides a student with the new tool to define a redundancy of a frequency dictionary.
Further, we shall understand the redundancy of a dictionary for the redundancy of a
sequence itself. There is a simpler way to define the redundancy; it is based on the
determination of the longest repeat within a sequence \cite{g3,g4,g5,g6}. It was found
that the redundancy of introns exceeds that latter for exons, and the splicing results in
a decrease of a general gene redundancy. Meanwhile, it should be stressed, that this
simple method of the redundancy determination fails to figure out the situations of
highly ordered (e.g. periodical) sequences. The definition of redundancy through an
information capacity calculation is free from that discrepancy. It should be kept in
mind, that zero value of (\ref{infcap}) does not automatically yield an exact and
unambiguous reconstruction of a thicker finite dictionary $\mathfrak{F}_q$. A redundancy,
then, is to be considered in two interrelated but individual ways; the former is the
measure defined through the $\mathfrak{F}_{d^{\ast}}$ perfect expansion up to an entire
symbol sequence, and the latter is a high level of predictability of a continuation of
each word.

\subsection{Markov models and maximum entropy principle}\label{mark}
Study of nucleotide sequences with the Markovian processes is rather popular
\cite{durak,borhy}. The motivation behind such popularity is quite transparent: Markov
process provides a researcher with numerous ways to fit a specific realization of the
process to a given symbol sequence. Basic idea of a search of so called hidden Markov
models of a nucleotide sequence consists in a choice of the minimal order Markov process,
which matches the sequence satisfactory. It should be said, that this approach
distinguishes quite properly coding regions of a genome vs. the non-coding ones
\cite{borhy}. The invariance in triplet distribution found for non-coding regions
accompanied with a distinct and well structured pattern of a triplet distribution
observed within coding regions \cite{z1,z1-1,z2} makes such good efficiency of Markov
models for separation of coding vs. non-coding regions evident.

Formally speaking, there always exists a Markov process, that perfectly fits a sequence.
Indeed, Markov process of the $d^{\ast}$ order developed over a sequence would match this
latter perfectly, with no variations, at all. Obviously, such Markov model brings no
biological inspiration. Nevertheless, a search for minimal order Markov process matching
a sequence may make sense. The key idea of a separation of coding regions from the
non-coding ones due to Markov models consists in a seeking for the points of an abrupt
change in the order of the relevant process. More fine and effective approach furthering
the hidden Markov model is discussed in Section~\ref{izs}.

\subsection{\label{frakt}On a fractal structure of genomes}
Comprehensive investigations of statistical properties of nucleotide sequences reveal
some interesting (and important) features of that latter. Researchers identify various
fractal structures and fractal-like patterns within genetic entities \cite{fr1,fr2}.
Probably, the nucleotide sequences are quite complex object exhibiting a great variety of
properties, including those, which are suspected to be a fractal pattern. A study of
information capacity reveals an increased correlation in a combinations of various
strings through the non-monotonic behaviour of that latter observed at different length
$q$ of words.

Examination of the tables shows a presence of the genomes that exhibit one, two or three
local minima of information capacity~(\ref{infcap}). Local minimum observed at the length
$l$ means that some combinations of two words of that length prevail among the others.
Correlations in short strings occurrence is evident, if several minima of information
capacity are observed. The correlations among some short strings are less evident, if a
single minimum is observed within a sequence. The point is that such single minimum may
result from a finite sampling of a sequence.

Nonetheless, one hardly could explain an occurrence of a single minimum of information
capacity~(\ref{infcap}) by a finite sampling effect, solely. The point is, that the
location of such singe minimum varies significantly, for various bacterial genomes.
Suppose, the location of the single minimum of information capacity is determined by the
finite sampling effect; then, the specific length $q$ of words where the minimum is
observed should be the same, for all such genomes. This follows from the dependence of
the maximum of~(\ref{infcap}); that former is defined mainly by the finite sampling
effect. Basically, the finite sampling effect would manifest through the logarithmic
dependence of the position of the maximum (and the minimum, in turn) on the length of a
sequence. Observed diversity of the lengths where the minimum occurs breaks down the
original supposition. Thus, the local minima (with no respect to the number of these
latter observed within a sequence) represent a structure, which might be considered as a
fractal pattern; detailed discussion of that matter falls beyond the scope of the paper.

\subsection{\label{izs}Information valuable words}
Let's have a look at the definition of information capacity~(\ref{vs}). It is evident,
that the major contribution into the sum is provided by the terms with the highest
possible deviation of real frequency $f_{\omega}$ from the most expected one
$\widetilde{f}_{\omega}$. These are the words of increased information value. More
exactly, $\alpha$, $\alpha>1$ be the information value threshold. A word $\omega'$ is of
information value, if it falls out of the range determined by the double inequality
\begin{equation}\label{alf}
\alpha^{-1} \leq \frac{f_{\omega'}}{\widetilde{f}_{\omega'}} \leq \alpha \;.
\end{equation}

There are two types of information valuable words: the former are the words with an
excess of real frequency over the expected one, and the latter are the words with an
excess of expected frequency over the real one. We call the words of the first type (of
the second type, respectively) the ascending ones (the descending ones, respectively).
Whether a word $\omega$ is of information value, or not, depends on a structure of a
sequence, of the threshold $\alpha$, and on the length $q$ of a word.

Of course, the choice of $\alpha$ value still is the matter of expertise of a student.
There is no formal way to put on the $\alpha$ level. To clarify this point, one has to
study the distribution of the words at a dictionary $W_q$ over their information value $p
= \frac{f_{\omega}}{\widetilde{f}_{\omega}}$. While the expected frequency
$\widetilde{f}_{\omega}$ is explicitly derived from the real frequency of the words
(see~(\ref{vosst})), less in known towards the distribution of words over the real (and
expected, in turn) frequency.

Obviously, $p_{\max}$ and $p_{\min}$ depend on a structure of nucleotide sequence. An
estimation for $p_{\min}$ is apparent: $\min{p_{\min}} = 0$; less is known concerning the
estimation of $p_{\max}$. To clarify this point, more studies should be carried out; they
fall beyond the scope of this paper.

Suppose, the threshold value $\alpha$ is put on. The threshold identifies two sets of
information valuable words; the former is the ascending one, and the latter is descending
one. A quality of being the information valuable word has no monotony: given ascending
information valuable word $\overline{\omega}$ (descending word $\underline{\omega}$,
respectively) of the length $q$ may be embedded into a longer one, or may be not.
Moreover, if no embedment is found for one symbol longer information valuable words, one
can not guarantee the embedment absence into the information valuable words of the length
$l$, where $l> q+1$. Besides, a longer information valuable word may incorporate a
shorter one with the opposite order of $p$.

Consider the uniform sets of information valuable words of increasing length:
$$\{\overline{\omega}_3\}, \{\overline{\omega}_4\}, \ldots, \{\overline{\omega}_q\} \quad
\textrm{and} \quad \{\underline{\omega}_3\}, \{\underline{\omega}_4\}, \ldots,
\{\underline{\omega}_q\}\;,$$ identified for given $\alpha>1$. A chain
\begin{subequations}\label{chain}
\begin{equation}\label{chain:up}
\overline{\omega}_3 \subset \overline{\omega}_4 \subset \ldots \subset
\overline{\omega}_k
\end{equation}
\textrm{or}
\begin{equation}\label{chain:down}
\underline{\omega}_3 \subset \underline{\omega}_4 \subset \ldots \subset
\underline{\omega}_k
\end{equation}
\end{subequations}
is the ascending shoot or descending shoot, respectively. The shortest word within a
shoot~(\ref{chain}) is a \textbf{root}, and the longest one is an \textbf{apex}. A union
of all the shoots with the same root makes a \textbf{pyramid}. Thus, a pyramid may be an
ascending, or a descending one. An information valuable word is an entity, where given
Markov model changes for another one (of the other order, etc.). A pyramid gathers the
entities where the variation of the relevant Markov model takes place in coordination,
for all the scales $3 \leq q \leq k$. Obviously, a simultaneous change of a Markov model
could hardly take place occasionally. Hence, the apices of the pyramids~(\ref{chain})
identifies the sites within a genome.

A study of distribution of the apices alongside a genome \cite{kitai,dan} shows a high
level of the correlation between a location of the apices, and the functional role of the
sites, where they occur. Such study makes a core of very promising approach in the
investigations of the relation of structure and function of biological macromoleculae,
while the detail discussion of that subject falls beyond the scope of this paper.

\begin{acknowledgments}
I would like to extend my gratitude to Prof.Alexander~N.Gorban from the University of
Leicester for long-time collaboration and permanent encouraging interest to the work, and
to Dr.Tatyana~G.Popova who draw my attention to the non-monotonicity of the information
capacity variation. I would like also to thank Marina~A.Makarova for help with
calculations.
\end{acknowledgments}


\begin{thebibliography}{99}

\bibitem{r1} W.-H.Li, \textit{Molecular Evolution} (Sineauer Associates, Sunderland, MA,
1997).

\bibitem{r2} \textit{Computational Models in Molecular Biology}, edited by S.L. Sazberg,
D.B. Searls, and S. Kasif (Elsevier, Amsterdam, 1998).

\bibitem{r3} J.K.Percus, \textit{Mathematics of Genome Analysis} (Columbia University
Press, Cambridge, 2002).

\bibitem{r4} A.K.Konopka, All we need is truth. // \textit{Comput.Biol.Chem}. (2004)
\textbf{28}(1), 1--2.

\bibitem{z1}A.Yu.Zinovjev, A.N.Gorban, T.G.Popova, Seven clusters in genomic triplet
distribution. \textit{In Siliko Biology} (2003) \textbf{3}, 471--482.

\bibitem{z1-1} A.N.Gorban, A.Yu.Zinovjev, T.G.Popova, Self-organizing approach
for automated gene identification. // \textit{Open Systems \& Information Dyn}. 2003
\textbf{10}, 321--333.

\bibitem{z2} A.Carbone, A.Zinovyev, F.Kepes, Codon Adaptation Index as a
measure of dominating codon bias. // \textit{Bioinformatics}. (2203) \textbf{19},
2005--2015.

\bibitem{z3} S.Havlin, S.V.Buldyrev, A.L.Goldberger, R.N.Mantegna, C.K.Peng,
M.Simons, H.E.Stanley, Statistical and linguistic features of DNA sequences. //
\textit{Fractals} (1995) \textbf{3}, 269--284.

\bibitem{z4} L.Allison, L.Stern, T.Edgoose, T.I.Dix, Sequence complexity for
biological sequence analysis. // \textit{Comput.Chem.} (2000) \textbf{24}(1), 43--55.

\bibitem{z5} V.N.Babenko, P.S.Kosarev, O.V.Vishnevsky, V.G.Levitsky, V.V.Basin,
A.S.Frolov, Investigating extended regulatory regions of genomic DNA sequences.
// \textit{Bioinformatics} (1999) \textbf{15}, 644--653.

\bibitem{z6} V.D.Gusev, L.A.Nemytikova, N.A.Chuzhanova, On the complexity
measures of genetic sequences. // \textit{Bioinformatics} (1999) \textbf{15}, 994--999.

\bibitem{a1} E.Pizzi, C.Frontali, Low-Complexity Regions in Plasmodium
falciparum Proteins. // \textit{Genome Res}. (2001) \textbf{11}, 218--229.

\bibitem{a2} H.E.Stanley, S.V.Buldyrev, A.L.Goldberger, S.Havlin, C.K.Peng, M.Simons,
Scaling features of noncoding DNA. // \textit{Physica}~\textbf{A} (1999) \textbf{273},
1--18.

\bibitem{a3} P.Romero, Z.Obradovic, X.Li, E.C.Garner, C.J.Brown, A.K.Dunker, Sequence
complexity of disordered protein. // \textit{Proteins}. (2001) \textbf{42}(1), 38--48.

\bibitem{a4} M.A.Jim\'{e}nez-Monta\~{n}o, W.Ebeling, Th.Pohl, P.E.Rapp,
Entropy and complexity of finite sequences as fluctuating quantities //
\textit{BioSystems} (2002) \textbf{64}, 23--32.

\bibitem{s4} E.O.Gorbunova, Yu.V.Kondratenko, M.G. Sadovsky, Data loss reparation due to
indeterminate fine-grained parallel computation. // P.M.A.Sloot et al. (Eds.) ICCS 2003,
LNCS 2658, Springer-Verlag, Berlin Heidelberg, 2003, pp.794--801.

\bibitem{s5} E.O.Nemechinskaya, Yu.V.Kondratenko, M.G. Sadovsky, Entropy based
approach to data loss reparation through the indeterminate fine-grained parallel
computation. // \textit{Open Systems \& Information Dyn}. (2004) \textbf{11}(2),
pp.161--175.

\bibitem{n1} N.N.Bugaenko, A.N.Gorban, M.G.Sadovsky, Towards the definition of
information content of nucleotide sequences. // \textit{Molecular biology Moscow}. (1996)
\textbf{30}, \#~5, 529--541.

\bibitem{n2} N.N.Bugaenko, A.N.Gorban, M.G.Sadovsky, The information capacity of
nucleotide sequences and their fragments. // \textit{Biophysics}. (1997) \textbf{5},
1063--1069.

\bibitem{n3} N.N.Bugaenko, A.N.Gorban, M.G.Sadovsky, Maximum entropy method in analysis
of genetic text and measurement of its information content. // \textit{Open Systems \&
Information Dyn}. (1998) \textbf{5}, \#~2, 265--278.

\bibitem{kitai} Rui~Hu, Bin~Wanga, Statistically significant strings are related to
regulatory elements in the promoter regions of {\sl Saccharomyces cerevisiae}. //
\textit{Physica}~\textbf{A} (2001) \textbf{290}, 464--474.

\bibitem{kirk} J.Kirkwood, E.Boggs, The radial distribution function in liquids. // \textit{J.Chem.
Physics} (1942) \textbf{10}, \#~6, p.394.

\bibitem{durak} M.S.Gelfand, Prediction of function in DNA sequence analysis. //
\textit{J.Comput.Biology} (1995) \textbf{2}, pp.87--115.

\bibitem{bal} R.Bales\c{c}u, \textit{Balance and imbalance statistical physics} (Sineauer Associates, Sunderland, MA,
1997).

\bibitem{obhod} A.N.Gorban, \textit{Equilibrium encircling}. Nauka plc.: 1984,
Novosibirsk. 268 p.

\bibitem{red1} M.J.Berryman, A.Allison, D.Abbott, Mutual information for examining correlations in
DNA. // arXiv:q-bio.PE/0404010v1; 7 April, 2004.

\bibitem{red2} Bailin~Hao, Huimin~Xie, Zuguo~Yu, Guoyi~Chen, Avoided Strings in Bacterial
Complete Genomes and a Related Combinatorial Problem. // \textit{Annals of Combinatorics}
(2000) \textbf{4}, 247--255.

\bibitem{red3} J.G.Wolff, Mathematics and logic as information compression by multiple
alignment, unification and search. // arXiv:math.GM/0308153v1 15 Augest, 2003.

\bibitem{red4} M.Li, P.Vit\'{a}nyi, \textit{An Introduction to Kolmogorov Complexity and
Its Applications}. Springer-Verlag: 1997, New York.

\bibitem{red5} C.E.Shannon, W.Weaver, \textit{The Mathematical Theory of Communication}.
University of Illinois Press: 1949, Urbana.

\bibitem{g3} T.G.Popova, M.G.Sadovsky, Splicing results in decrease of gene redundancy.
// \textit{Molecular Biology Moscow} (1995) \textbf{29}(3), 500--506.

\bibitem{g4} T.G.Popova, M.G.Sadovsky, Introns differ from exons in their redundancy. //
\textit{Rus.J.of Genetics} (1995) \textbf{31}(10), 1365--1369.

\bibitem{g5} A.N.Gorban, T.G.Popova, M.G.Sadovsky, Human viruses genes are less redundant
than the human genes. // \textit{Rus.J.of Genetics} (1996) \textbf{32}(2), 281--294.

\bibitem{g6} M.G.Sadovsky, On the redundancy of viral and prokaryotic genomes. //
\textit{Rus.J.of Genetics} (2002) \textbf{38}(5), 695--701.

\bibitem{kg1} A.N.Gorban, I.V.Karlin, Method of invariant manifolds and regularization of
acoustic spectra. // \textit{Transport Theory and Stat.Phys}. (1994) \textbf{23},
559--632.

\bibitem{kg2} A.N.Gorban, I.V.Karlin, General approach to constructing models of the Boltzmann
equation. // \textit{Physica}~\textbf{A} (1994) \textbf{206}, 401--420.

\bibitem{kg3} I.V.Karlin, A.N.Gorban, G.Dukek, T.F.Nonnenmacher, Dynamic correction to
moment approximations. // \textit{Phys.Rev.}~\textbf{E} (1998) \textbf{57}, 1668–-1672.

\bibitem{kg4} A.N.Gorban, I.V.Karlin, V.B.Zmievskii, S.V.Dymova, Reduced description in
reaction kinetics. // \textit{Physica}~\textbf{A} (2000) \textbf{275}(3-4), 361--379.

\bibitem{kg5} A.N.Gorban, I.V.Karlin, Uniqueness of thermodynamic projector and kinetic basis of molecular individualism. //
\textit{Physica}~\textbf{A} (2004) \textbf{336}(3-4), 391--432.

\bibitem{kg6} A.N.Gorban, I.V.Karlin, Family of additive entropy functions out of
thermodynamic limit. // \textit{Phys.Rev.}~\textbf{E} (2003) \textbf{67}, 016104.

\bibitem{witten2001} T.D.Schneider, Evolution of biological information. // \textit{Nucleic
Acids Res}. (2000) \textbf{28}, 2794--2799.

\bibitem{borhy} W.S.Hyes, M.Borodovsky, How to interpret anonimous bacterial genome:
Machine Learning Approach to Gene Identification. // \textit{Genome Res}. (20??)
\textbf{??}, 2794--2799.

\bibitem{fr1} D.R.Bickel, B.J.West, Multiplicative and fractal processes in DNA
evolution. \textit{Fractals} (1998) \textbf{6}, 211--217.

\bibitem{fr2} N.N.Oiwa, J.A.Glazier, The fractal structure of the
mitochondrial genomes. // \textit{Physica}~\textbf{A} (2002) \textbf{311}(3-4), 221--230.

\bibitem{dan} M.A.Makarova, M.G.Sadovsky, Information approach in the problem of
structure--function relation in biological macromoleculae. // Doklady Biochemistry and
Biophysics (2004) \textbf{6}, \#~1, 236--240.

\end{thebibliography}
\end{document}